\documentclass[12pt]{article}
\usepackage{epsf}

\usepackage{epsfig,graphics}
\usepackage {graphicx}
\usepackage {epsfig}
\usepackage {subfigure}
\usepackage {tabularx} 
\usepackage{rotate}	
\usepackage{slashed}

\usepackage{amsmath}
\usepackage{amsfonts}
\usepackage{amssymb}
\usepackage{graphicx}
\usepackage{cite}

\newcommand{\bmat}{\left(\begin{array}}
\newcommand{\emat}{\end{array}\right)}

\def\yzero{\smash{\hbox{$y\kern-4pt\raise1pt\hbox{${}^\circ$}$}}}

\def\a{\alpha}

\def\beq{\begin{equation}}
\def\eeq{\end{equation}}
\def\beqa{\begin{eqnarray}}
\def\eeqa{\end{eqnarray}}

\def\-{\hphantom{-}}

\def\s2{\frac{1}{\sqrt2}}

\def\beq{\begin{equation}}
\def\eeq{\end{equation}}
\def\beqa{\begin{eqnarray}}
\def\eeqa{\end{eqnarray}}

\def\IF{\relax{\rm I\kern-.18em F}}
\def\II{\relax{\rm I\kern-.18em I}}
\def\IP{\relax{\rm I\kern-.18em P}}
\def\IC{\relax\hbox{\kern.25em$\inbar\kern-.3em{\rm C}$}}
\def\IR{\relax{\rm I\kern-.18em R}}

\def\Dsl{\,\raise.15ex\hbox{/}\mkern-13.5mu D} %this one can be subscripted
\def\IZ{Z\kern-.4em  Z}

\catcode`\@=11   
\newdimen\@rotdimen
\newbox\@rotbox  

\def\@vspec#1{\special{ps:#1}}%  passes #1 verbatim to the output
\def\@rotstart#1{\@vspec{gsave currentpoint currentpoint translate
   #1 neg exch neg exch translate}}% #1 can be any origin-fixing transformation
\def\@rotfinish{\@vspec{currentpoint grestore moveto}}% gets back in synch 
\def\@rotr#1{\@rotdimen=\ht#1\advance\@rotdimen by\dp#1%
   \hbox to\@rotdimen{\hskip\ht#1\vbox to\wd#1{\@rotstart{90 rotate}%
   \box#1\vss}\hss}\@rotfinish}
\def\@rotl#1{\@rotdimen=\ht#1\advance\@rotdimen by\dp#1%
   \hbox to\@rotdimen{\vbox to\wd#1{\vskip\wd#1\@rotstart{270 rotate}%
   \box#1\vss}\hss}\@rotfinish}%
\def\@rotu#1{\@rotdimen=\ht#1\advance\@rotdimen by\dp#1%
   \hbox to\wd#1{\hskip\wd#1\vbox to\@rotdimen{\vskip\@rotdimen
   \@rotstart{-1 dup scale}\box#1\vss}\hss}\@rotfinish}%
\def\@rotf#1{\hbox to\wd#1{\hskip\wd#1\@rotstart{-1 1 scale}%
   \box#1\hss}\@rotfinish}%
\def\rotate{\@ifnextchar[{\@rotate}{\@rotate[l]}}
\def\@rotate[#1]#2{\setbox\@rotbox=\hbox{#2}\@nameuse{@rot#1}\@rotbox}

\catcode`\@=12
\topmargin
-1.5cm
\textwidth
15.5cm
\textheight
23.5cm
\oddsidemargin
0.7cm
\evensidemargin
0.7cm

\begin{document}

%----------------------------------------------------------------------%
%  numbering equations with section number
%----------------------------------------------------------------------%
\makeatletter
\@addtoreset{equation}{section}
\makeatother
\renewcommand{\theequation}{\thesection.\arabic{equation}}
%----------------------------------------------------------------------%
%  title page
%----------------------------------------------------------------------%
\pagestyle{empty}
%\vspace*{1.0in}
\vspace{-0.2cm}
\rightline{IFT-UAM/CSIC-15-141}
\rightline{FTUAM-15-45}
%\rightline{\tt hep-th/xxxxxxx}
\vspace{1.2cm}
\begin{center}

%\vspace{0.5cm}

\LARGE{ A Megaxion at 750 GeV as a First Hint \\ of Low Scale String Theory} 
%({\it Questo e una genialit\'a o e tutto una cassata})
\\[13mm]
  \large{ Luis E. Ib\'a\~nez$  $ and V\'ictor Mart\'in-Lozano   \\[6mm]}
\small{
  Departamento de F\'{\i}sica Te\'orica
and Instituto de F\'{\i}sica Te\'orica UAM/CSIC,\\[-0.3em]
Universidad Aut\'onoma de Madrid,
Cantoblanco, 28049 Madrid, Spain 
\\[8mm]}
\small{\bf Abstract} \\[7mm]
\end{center}
\begin{center}
\begin{minipage}[h]{15.22cm}
Low scale string models naturally have axion-like pseudoscalars which couple directly to gluons and
photons (but not $W$'s) at tree level.  We show how they typically get tree level masses in the presence of  closed
string fluxes , consistent with the axion discrete  gauge symmetry, in a way akin of
the axion monodromy of string inflation and relaxion models. We discuss the possibility that the hints for a resonance at 750 GeV recently reported at
ATLAS and CMS  could correspond to such a  heavy axion state ({\it megaxion}).  Adjusting the production rate and branching ratios 
suggest  the string scale to be of order $M_s\simeq 7-10^4$ TeV, depending on the compactification geometry.
If this interpretation was correct, one extra $Z$' gauge boson  could be produced before reaching the
string threshold at LHC and  future colliders.

\end{minipage}
\end{center}
\newpage
%----------------------------------------------------------------------%
%  Resetting of counters
%----------------------------------------------------------------------%
\setcounter{page}{1}
\pagestyle{plain}
\renewcommand{\thefootnote}{\arabic{footnote}}
\setcounter{footnote}{0}
%----------------------------------------------------------------------%
%  Paper begins
%----------------------------------------------------------------------%

%&&&&&&&&&&&&&&&&&&&&&&&&&&&&&&&&&
\section{Introduction}
%&&&&&&&&&&&&&&&&&&&&&&&&&&&&&&&&&

ATLAS and CMS collaborations have recently reported hints of a 750 GeV  resonance in the diphoton search \cite{diphoton:atlas,diphoton:cms} for integrated luminosities of 3.2 fb$^{-1}$ and 2.6 fb$^{-1}$ respectively. The ATLAS collaboration presents an excess with a local significance of 3.6 $\sigma$ while CMS collaboration obtained a local significance of 2.6 $\sigma$. These two statistical significances correspond to cross sections of $\sigma (pp\to\gamma\gamma)\sim 10.6$ fb (ATLAS) and $\sigma(pp\to\gamma\gamma)\sim 6.3$ fb (CMS). It is useful to compare these results to  the diphoton searches in the first  LHC run. The CMS diphoton search for a centre-of mass energy of $\sqrt{s}=8$ TeV for an integrated luminosity of 19.7 fb$^{-1}$ reports a mild excess for a mass of 750 GeV with a cross section of $\sigma(pp\to\gamma\gamma)\sim 0.5$ fb \cite{diphoton:cms8tev}, while ATLAS collaboration obtained $\sigma(pp\to\gamma\gamma)\sim 0.4$ fb for the same mass \cite{Aad:2015mna}. The 13 TeV data from ATLAS indicate a preferred value for the resonance width of $\Gamma=45$ GeV \cite{diphoton:atlas} that supposes a not-so narrow width ($\Gamma/M \sim 6\%$). On the other hand CMS results suggest a better agreement with a narrow width, however when fitting the data they show that a width of $\Gamma=42$ GeV is also compatible \cite{diphoton:cms}. In that sense we can estimate that the resonance has an upper limit on its width of $\Gamma \lesssim 45$ GeV. 

One can also interpret this resonance as a particle and obtain information about it from the different experimental data. In that sense a pseudoscalar particle is highly motivated from the results reported by ATLAS and CMS. First of all the only allowed spins for a resonant particle decaying into two photons are 0 and 2 by the Landau-Yang theorem. Moreover if we compare the ratio between the cross sections at $\sqrt{s}=8$ TeV and 13 TeV we obtain a factor of 5, this coincides with the gain factor of the production cross section of a (pseudo) scalar particle produced by gluons for those energies at a mass of 750 GeV \cite{LHCcrosssection}.  However different searches at $\sqrt{s}=8$ TeV present null results in searching for resonant production of particles decaying into Standard Model (SM) final states such as $t\bar{t}$\cite{Chatrchyan:2013lca}, $WW$\cite{Aad:2015agg, Khachatryan:2015cwa}, $ZZ$\cite{Aad:2015kna, Khachatryan:2015cwa}, $Z\gamma$\cite{Aad:2014fha}, $\ell^+\ell^-$\cite{Aad:2014cka, Khachatryan:2014fba}, $b\bar{b}$\cite{Khachatryan:2015tra}... Data seem to indicate that the pseudoscalar resonant particle only couples to gluons and photons. Several papers trying to disentangle the diphoton resonance in terms of axions or other different models could be found in refs.~\cite{popurri,Franceschini:2015kwy,madrugadores}.

The effective Lagrangian of  an  axion  $\eta$ coupled to gluons and photons is
\begin{equation}
\mathcal{L}_{a_0}=\frac{\alpha_s}{4\pi} g_{g}\eta G_{\mu\nu}\widetilde{G}^{\mu\nu} + \frac{\alpha_{em}}{4\pi} g_{\gamma}\eta F^{em}_{\mu\nu}\widetilde{F}_{em}^{\mu\nu},
\label{eq:Lag}
\end{equation}
where $\alpha_s$ and $\alpha_{em}$ are the strong and electromagnetic fine-structure constants, $g_g$ and $g_\gamma$ are  model dependent 
constants.  Axions $\eta_i $ are related to their canonically normalised axions $a_i$ by  $\eta_i =a_i/f_i$, so that the kinetic term is 
\beq
L_{f} \ =\   -\frac {1}{2} \partial_\mu a_i\partial^\mu a_i \ =\ -\frac {f_i^2}{2} \partial_\mu \eta_i \partial^\mu \eta_i \ .
\eeq
and $f_i$ are  the corresponding axion decay constants.
As we will see, in order to match the data hints with an axion we will need parameters in the range  $f/g_g\simeq 10^2-10^3$ GeV, $f/g_\gamma\simeq 1-10^2$ GeV.
We also need an explanation as to why the axion does not couple to $W$'s but it does couple to gluons and photons.
 And finally, we need an explanation as to how an axion-like
object is so heavy, of order 750 GeV. Usual axions in particle physics are perturbatively massless and only acquire a mass due to non-perturbative 
potential. This potential is generated by instantons and is periodic under shifts $a_0\rightarrow a_0+2\pi f$, which is an unbroken discrete 
gauge symmetry, and is the characteristic feature of an axion-like field.

It has been realised in the last few years that
axion-like objects can get a perturbative mass term and still preserve the discrete shift symmetry if at the same time the parameters in the potential shift appropriately. 
These type of axions are sometimes called {\it monodromy axions}\cite{mono,KS,msu,BIV}  and the simplest implementation of its symmetries is in terms of Minkowski 3-form fields
$C_{\mu \nu \rho}$. Such 3-form fields do not propagate, since the corresponding equations of motion fix its field-strength to be
constant, $F_{\mu \nu \rho \sigma}=\epsilon_{\mu \nu \rho \sigma}f_0$ \cite{KS,msu,BIV}. The required structure is obtained from the following action \cite{KS}
\beq
L=-\frac12(\partial_{\mu} a_0)^2-\frac12|F_4|^2+\mu a_0F_4,
\label{KSpot}
\eeq
where 
$F_4=\epsilon^{\mu \nu \rho \sigma}F_{\mu \nu \rho \sigma}$. Since the 3-form field has no propagating degrees of freedom in 4d, it behaves like an auxiliary field.
Its equation of motion yields
\beq
F_4=f_0\ +\ \mu a_0
\eeq
leading to an induced scalar potential for the axion
\beq
V_a\ =\ \frac12(f_0\ +\ \mu a_0)^2.
\label{V0}
\eeq
This potential is invariant under the combined shift
\beq
a_0\ \rightarrow \ a_0 \ +\ 2\pi f  \ ;\    f_0\ \rightarrow \ f_0 \ -\ 2\pi \mu f  \ .
\eeq
As noted in \cite{BP} the 4-form  vev $f_0$ is quantized,  and consistency with the symmetries requires $2\pi \mu f$ to be an integer in the same mass$^2$ units as
$f_0$. Note that the axion $a_0$ has a mass $m_a=\mu$ and still the  discrete shift symmetry is maintained.  This class of
monodromy axions have been recently considered in the context of string monodromy inflation  \cite{mono,KS,msu}
and more recently in the context of relaxion 
dynamics \cite{relaxion}.

The structure of the rest of the paper is as follows. In the next section we will show how a heavy axion with the couplings discussed
above is compatible with the hints of a 750 GeV boson observed at CMS and ATLAS. In section (\ref{stringy}) we will show how 
all the required ingredients are  simultaneously present in string theory models with a string 
scale in the range  $7-10^4$ TeV.  Once this work was finished refs.\cite{madrugadores} appeared  (a few days or hours before our submission)
which also consider the possibility  of a string axion-like being the 750 GeV state.

\section{A Megaxion and the 750 GeV Excess}

A simple analysis of the cross section reported by ATLAS and CMS for 13 TeV \cite{diphoton:atlas, diphoton:cms} gives us a central value of $\sigma_{\gamma\gamma}=7.6\pm 1.9$ fb, we will take this value in the rest  of the paper. 
The production cross section of the axion decaying into two photons can be written as \cite{Franceschini:2015kwy}
\begin{equation}
\sigma(pp\to a_0\to \gamma\gamma)= \frac{C_{gg}}{\Gamma_{a_0}m_{a_0}s}\Gamma(a_0\to gg)\Gamma(a_0\to \gamma\gamma),
\label{eq:diphoton}
\end{equation}
where we have used the narrow width approximation (NWA).\footnote{We have assumed the NWA to obtain eq.(\ref{eq:diphoton}). However as ATLAS collaboration reports the width of the resonance is compatible with a value of $\Gamma=45$ GeV. In that case the error of taking this approximation is of the order of $\mathcal{O}(\Gamma/M)\sim 6 \%$.} $C_{gg}$ is the partonic integral for gluon production of the pseudoscalar,  whose value for 13 TeV is $C_{gg}^{13{\rm TeV}}=2137$ \cite{LHCcrosssection}. The decay widths of the axion decaying into gluons and photons are
\begin{eqnarray}
\Gamma(a_0\to \gamma\gamma) = \kappa_\gamma^2\frac{m_{a_0}^3}{64\pi},  \label{eq:decayga}\\
\Gamma(a_0\to gg)= \kappa_g^2\frac{m_{a_0}^3}{8\pi}.\label{eq:decayg}
\end{eqnarray}
Here we have defined
\begin{equation}
\kappa_i=\frac{\alpha_i}{4\pi}\frac{g_i}{f},\quad i=g,\gamma .
\end{equation}
Note that with the defining Lagrangian 
eq.(\ref{eq:Lag}),   one has  $\Gamma_{a_0}=\Gamma(a_0\to gg) + \Gamma(a_0\to \gamma\gamma)$, since no other decays are possible to leading order.

The axion can also decay into gluons in such a way that the dijet searches could be sensitive to it. The results of this search at $\sqrt{s}=8$ TeV performed by ATLAS \cite{Aad:2014aqa}  and CMS \cite{dijet:cms8tev} lead to an upper bound on the cross section of $\sigma_{jj}\lesssim 2.5$ pb for a mass of 750 GeV. In our case the dijet cross section is given by
\begin{equation}
\sigma(pp\to a_0\to jj)= \frac{C_{gg}}{\Gamma_{a_0}m_{a_0}s}\Gamma(a_0\to gg)^2.
\label{eq:dijet}
\end{equation}
It is clear from eq.(\ref{eq:dijet}) that this bound imposes a constant upper limit on $\kappa_g$ if $\kappa_\gamma \ll \kappa_g$ and it becomes weaker as long as $\kappa_\gamma$ grows.

\begin{figure*}
\begin{center}
\includegraphics[scale=0.43]{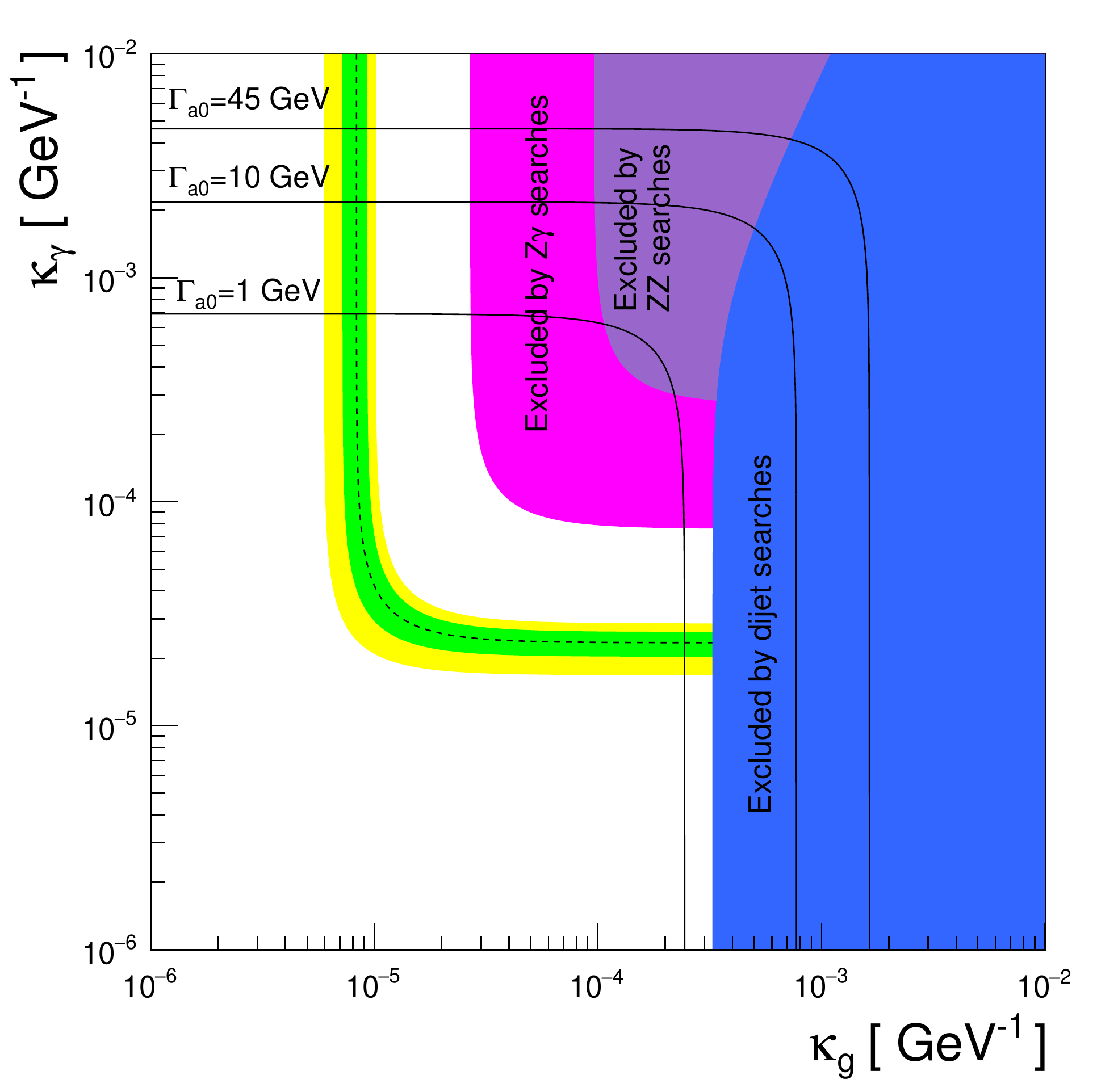}
\end{center}
\caption{\footnotesize The effective coupling of the axion to photons $\kappa_\gamma$ versus the effective coupling of the axion versus gluons $\kappa_g$. The central value of the cross section of the excess reported by ATLAS and CMS is shown as a black dashed line while the green and yellow bands indicate the 1$\sigma$ and 2$\sigma$ regions. The solid black lines represent different values of the axion decay width that are $\Gamma_{a_0}=45,\, 10,\, 1$ GeV. The blue area defines the region excluded by dijet searches \cite{Aad:2014aqa, dijet:cms8tev}, the violet and magenta areas are the regions excluded by $ZZ$ and $Z\gamma$ searches respectively \cite{Aad:2015kna,Khachatryan:2015cwa,Aad:2014fha}.}
      \label{fig:constraints}
\end{figure*}

\begin{figure*}[h!]
\begin{center}
\includegraphics[scale=0.43]{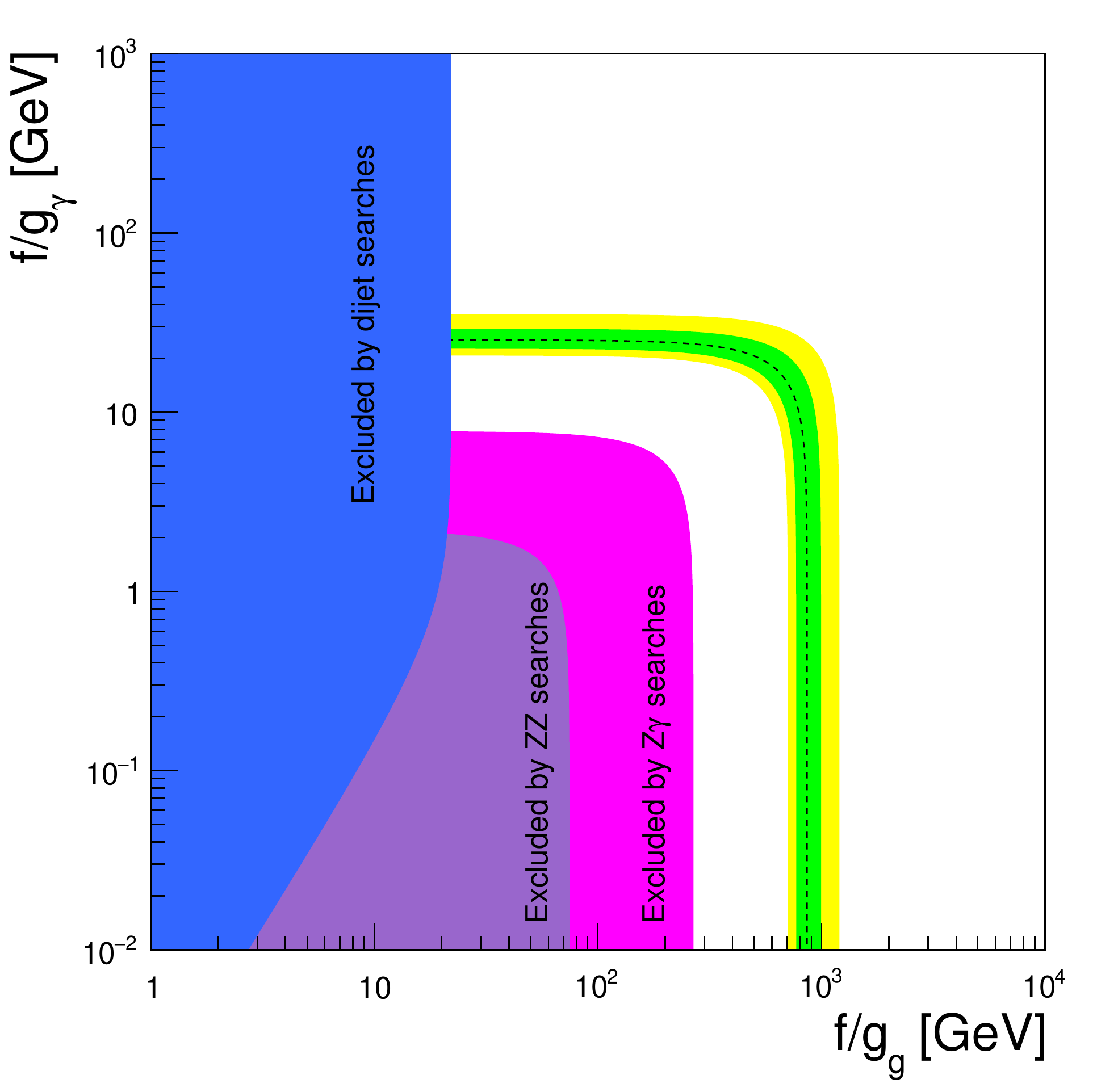}
\end{center}
\caption{\footnotesize $f/g_\gamma$ versus $f/g_g$. The central value of the cross section of the excess reported by ATLAS and CMS is shown as a black dashed line while the green and yellow bands indicate the 1$\sigma$ and 2$\sigma$ regions. The solid black lines represent different values of the axion decay width that are $\Gamma_{a_0}=45,\, 10,\, 1$ GeV. The blue area defines the region excluded by dijet searches \cite{Aad:2014aqa, dijet:cms8tev}, the violet and magenta areas are the regions excluded by $ZZ$ and $Z\gamma$ searches respectively \cite{Aad:2015kna,Khachatryan:2015cwa,Aad:2014fha}.}
      \label{fig:constraints2}
\end{figure*}

The results obtained are shown in fig.~(\ref{fig:constraints}) where we have plotted the diphoton cross section data in the plane ($\kappa_\gamma,\kappa_g$) as a black dashed line and the green and yellow bands indicate the 1$\sigma$ and 2$\sigma$ values for this cross section. The vertical line for the cross section is given by the fact that the minimal value of $\kappa_g$ to give the correct production of the axion is approximately of the order $10^{-5}$ GeV$^{-1}$ and it is constant for any bigger  value of $\kappa_\gamma$,  since in that limit the decay width of the axion is mainly the decay width into photons, $\Gamma_{a_0}\approx \Gamma (a_0\to \gamma\gamma)$. The horizontal line for the cross section can be understood in the same way as before changing $\kappa_g$ by $\kappa_\gamma$ and viceversa. The full black lines represent different values of $\kappa_g$ and $\kappa_\gamma$ giving decay widths for the axion of 45 GeV, 10 GeV and 1 GeV for illustration. We take the line of $\Gamma=45$ GeV as an upper limit on the decay width of the axion so the area above that line represents a greater value than the one given by ATLAS. The blue region represents the exclusion region given by dijet searches at $\sqrt{s}$=8 TeV \cite{Aad:2014aqa,dijet:cms8tev}. For low values of $\kappa_\gamma$ the exclusion limit is constant in $\kappa_g$ since for those values the decay width of the axion is dominated by the decay into gluons. However when $\kappa_\gamma$ grows to values greater than $10^{-3}$ GeV$^{-1}$ the decay into photons becomes important so the dijet bound weakens. If we want to explain the cross section given by the excess of 750 GeV, the dijet searches impose a bound on the maximum value of $\kappa_g$ of approximately $\kappa_g\lesssim 3\times 10^{-4}$ GeV. On the other hand the maximum value for $\kappa_\gamma$ is given by the maximum value of the axion decay width providing an upper limit of $\kappa_\gamma \lesssim 5\times 10^{-3}$ GeV$^{-1}$.  The preferred value for the diphoton cross section provide two possible windows, one of this windows is the one with $7\times 10^{-6}\, {\rm GeV}^{-1} \lesssim \kappa_g \lesssim 1.2\times 10^{-5}\, {\rm GeV}^{-1}$ that is valid for values of $\kappa_\gamma$ greater than $\kappa_\gamma \gtrsim 1.7\times 10^{-5}\, {\rm GeV}^{-1}$. The other window corresponds to values of $\kappa_\gamma$ that lie on the region $1.7\times 10^{-5}\, {\rm GeV}^{-1} \lesssim \kappa_g \lesssim 2.7\times 10^{-5}\, {\rm GeV}^{-1}$ for values of $\kappa_g$ that are greater than $\kappa_g\gtrsim 7\times 10^{-6}$ GeV$^{-1}$. Due to the exclusion given by dijet searches there is only a small region that can provide the total width of the axion to be $\Gamma_{a_0}=45$ GeV, this region is characterised by $\kappa_\gamma=4.7\,{\rm  GeV}^{-1}$ and $7\times 10^{-6}\, {\rm GeV}^{-1} \lesssim \kappa_g \lesssim 1.2\times 10^{-5}\, {\rm GeV}^{-1}$. We have also shown the areas constrained by $ZZ$ and $Z\gamma$ searches at 8 TeV \cite{Aad:2015kna,Khachatryan:2015cwa,Aad:2014fha}  as violet and magenta regions. It is clear from fig.~(\ref{fig:constraints}) that the bounds imposed by those searches do not affect the signal cross section of the 750 GeV resonance. 

In fig.~(\ref{fig:constraints2}) the diphoton cross section is depicted in the plane $(f/g_\gamma,f/g_g)$. As in the previous case the central value of the diphoton cross section is represented as a black dashed line and the 1$\sigma$ and 2$\sigma$ values are the green and yellow bands. The dijet exclusion area is represented as the blue region, and the $ZZ$ and $Z\gamma$ exclusion areas are shown as violet and magenta regions. From this figure we can obtain the values of $f/g_\gamma$ and $f/g_g$ that reproduce the diphoton cross section divided in two different windows. The first one is for values in the regions $20 \,{\rm GeV}\lesssim f/g_\gamma \lesssim 30\,{\rm GeV}$ and $20 \,{\rm GeV}\lesssim f/g_g \lesssim 10^3\,{\rm GeV}$, while the second one take the values from the regions $10^{-2} \,{\rm GeV}\lesssim f/g_\gamma \lesssim 30\,{\rm GeV}$ and $7\times 10^2 \,{\rm GeV}\lesssim f/g_g \lesssim 1.1\times 10^3\,{\rm GeV}$. If we consider that the values for $g_g$ and $g_\gamma$ are of order $\mathcal{O}(1)$ the values allowed for the axion decay constant are $1\,{\rm GeV}\lesssim f \lesssim 10^{3} {\rm GeV}$.
 
 In summary, an axion $a_0$ with couplings so constrained is consistent with the observed hints of a $750$ GeV boson. Such limits will in turn 
 constraint the structure of possible low scale string models whose structure we discuss next.

%It is important to remark that in this kind of construction the effective couplings of the axion to gluons and photons are not independent. They follow the following relation,
%\begin{equation}
%\frac{\kappa_g}{\kappa_\gamma}=a\frac{\alpha_s}{\alpha_{em}},
%\end{equation}
%where $a$ is an integer number and it is given by ...(stringy stuff) It is important to notice that in this relation we loose the information about the string scale contained in the %parameter $f$.

\section{A Megaxion at 750 GeV as a Hint of Low Scale String Theory} \label{stringy}

It is well known that the string theory scale may in principle be very low, even of order slightly above the EW scale, e.g. $M_s\simeq 7-10^4$ TeV
\cite{Antoniadis:1998ig} (see \cite{BOOK,interev} for reviews).
No sign of string resonances
have been observed yet at LHC, indicating  a lower bound for the string scale, e.g.  $M_s\geq 7$ TeV \cite{Khachatryan:2015dcf}. In this scenario the fact that the Planck scale is much
bigger than the string scale, $M_p\gg M_s$, is due to some extra dimensions  (transverse  to the branes in which the SM resides) being
very large. 

\begin{figure*}
\begin{center}
\includegraphics[scale=0.3]{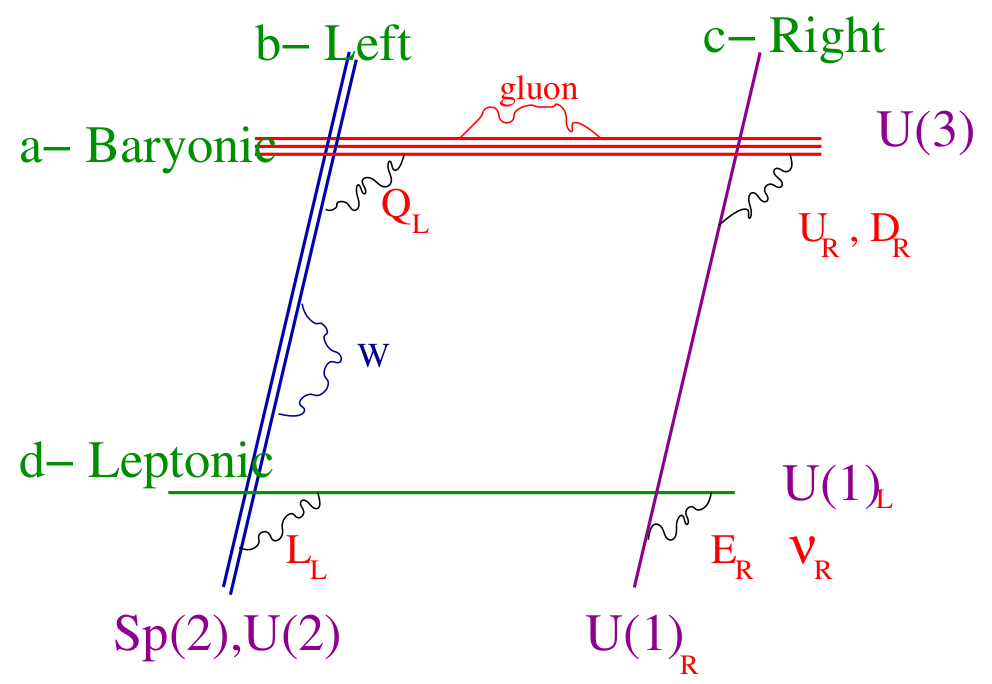}
\end{center}
\caption{\footnotesize Quarks and leptons at intersecting branes.}
      \label{fig:branes}
\end{figure*}

In what follows we will assume that the string scale is in the  mentioned TeV range. We will mostly use for illustrative purposes a particularly interesting class of 
string models  based on Type IIA orientifolds  with intersecting branes \cite{BOOK,interev}, see the Appendix.  In these models the observable fermionic sector is that of the SM \cite{IMR,IMRD5}.
The scheme of this large class of models is depicted in fig.~(\ref{fig:branes}) . The quarks and leptons reside  at the intersection of $D6$-branes which come in 
4 stacks labeled a,b,c,d, and leading to a gauge group $U(3)_a\times U(2)_b\times U(1)_c\times U(1)_d$ respectively (the EW group may also be $Sp(2)\approx  SU(2)$ if a single brane sits on top
of the orientifold plane).  The 4 stacks a,b,c,d of branes are called
baryonic, weak, right and leptonic, because  of the associated gauge symmetries.
In addition to $SU(3)\times SU(2)$, the (visible) gauge group has thus
up to 4  $U(1)$'s all of which get a mass of order the string scale \footnote{Or rather somewhat below, see \cite{Ghilencea:2002da,Ghilencea:2002by}  
and comments at the end of this section.}  by the Green-Schwarz 
mechanism except for hypercharge,  which is a linear combination of the 4 $U(1)$'s. In particular one has
\beq
Q_Y \ =\ \frac {1}{6}Q_a \ -\ \frac {1}{2} Q_c \ +\ \frac {1}{2}Q_d \ .
\label{hyper}
\eeq
In addition to the SM particles these models come along with scalar singlets coming from the closed string sector of the theory, the complex structure and Kahler 
moduli fields\cite{IMR}.  Among these there are always a set of axion-like fields coming from the Ramond-Ramond sector of the theory and in SUSY models become the imaginary part
of the complex structure fields, $Im\,U_i = a_i$.  They come from the dimensional reduction of  RR 3-forms $C_3$ with legs in
internal dimensions. In the toroidal setting there are 4 such scalars $i =0,1,2,3$\cite{IMR}.  As we said, some of these would be axions 
get mass by combining with three linear combinations of the $U(1)$'s in the theory.  To see how this happens it is more useful to consider an equivalent
description of these axions in terms of 2-forms $B^{\mu \nu}_i$. They are related to the pseudo scalars by 
$\epsilon_{\mu \nu \rho \sigma}\partial^\sigma a_i = H^i_{\mu \nu \rho}$, where $H=dB$ is the field strength of each 2-form. There are then couplings \cite{IMR}
\beq
c_i^\alpha \ B_i \wedge F_{U(1)}^\alpha  \ \ \   \  \alpha\ =\ a,b,c,d  \ ;  \  i=0,1,2,3 \ ,
\eeq
where $F_{U(1)}^\alpha $ are the field strengths of the 4 $U(1)$'s. The coefficients $c_i$ are
integers in an appropriate normalisation.  These couplings, when written in terms of the axions $a_i$ are Higgs-like 
couplings which render massive all of the $U(1)$'s except for hypercharge. In particular one has couplings (see the Appendix) 
\beqa
c_1^b  \ B_1 &\wedge & F^b  \\ \nonumber
c_2^d \  B_2 & \wedge & \ (-3F^a \ +\ F^d) \\ \nonumber
B_3 & \wedge & [c_3^a F^a \ +\ c_3^b F^b\ + \ (\frac {1}{3} c_3^a+c_3^d)F^c\ + \ c_3^dF^d)] \ .
\label{BFS}
\eeqa
Note the important point that the 2-form $B_0$ (or its corresponding dual, the axion $a_0$)  does not appear in any of
these couplings and hence it does not combine with any gauge boson and remains massless at this level.
This is more general than the toroidal setting. Generically there are axion fields like $a_0$ which have couplings to gauge bosons 
but are not the Goldstone boson of any $U(1)$.

In addition to these couplings, the axions $a_i$ have also axion-like couplings of the form
\beq
d_i^\alpha \  a_i ( tr\ F_\alpha \wedge F_\alpha)  \ \ \    ;   \  \alpha\ =\ a,b,c,d  \ ; \  i=0,1,2,3 \ .
\eeq
The $d_i$'s are coefficients which are integers in an appropriate normalisation.
Here $F_\alpha$ are the full $U(n)$ field strengths of the 4 stacks. These couplings, combined with those in eq.(\ref{BFS}) cancel all  the residual 
mixed $U(1)$ triangle anomalies of  the massive $U(1)$'s.  The massless axion $a_0$ has in general such couplings, with a general form
\beq
a_0 \ [ d_0^a \ F^a \wedge F^a \ + \ d_0^b \ F^b \wedge F^b \ +\ d_0^d \ F^d\wedge F^d ] \ .
\label{c0coup}
\eeq
As we said, the coefficients $d_0$ are model dependent integers. In particular, as explained in the Appendix, in a large class of models $d_0^b=0$ and the unique massless
axion  will couple only to $SU(3)$ and to hypercharge (via $U(1)_a$ and $U(1)_d$, which do couple to $a_0$).  In the class of toroidal models 
discussed in the Appendix this happens when one has an integer $n_b^1=0$.  Thus we are left with axion couplings of the general form discussed in
the introduction, i.e.
\begin{equation}
\mathcal{L}_{a_0}=\frac{\alpha_s}{4\pi} g_{g}\frac {a_0}{f} G_{\mu\nu}\widetilde{G}^{\mu\nu} +
 \frac{\alpha_{Y}}{4\pi} g_{Y} \frac {a_0}{f} B_{\mu\nu}\widetilde{B}^{\mu\nu},
\label{axioncoup}
\end{equation}
Note that the dependence on the couplings $\alpha_s,\alpha_Y$ arises once one sets the gauge kinetic terms 
$F^2/(4g^2)$ to canonical form, whereas   $g_g$,$g_Y$ are model dependent constant coefficients .
In the case of the QCD coupling $g_g$ is proportional to the $d_0^a$ coefficient. However, in the case of hypercharge 
it will be connected to  both $d_0^a$ and $d_0^d$, since both $U(1)^a$ and $U(1)^d$ appear in the definition of hypercharge,
see eq.(\ref{hyper}). Furthermore, in the case of hypercharge several branes are involved and the geometric factors $\xi$ (see below)
will in general affect differently the different branes. The upshot  is that {\it  $g_g/f$ and $g_Y/f$ should be considered independent
parameters}, to be fixed by experiment.  This is what we have done in the phenomenological analysis in the previous section.

In summary this class of Type IIA orientifold models generically has a single axion-like field $a_0$ which remains light after 
the $U(1)$'s other than hypercharge get a mass. Morover, there are large classes of models in which this axion has couplings to gluons
and hypercharge but not to $SU(2)$ gauge bosons.  It is remarkable that these conditions, required by experimental data, 
appear in the model so neatly. 

The size of the axion couplings to gluons and photons is controlled by the value of the axion decay constant $f$, which in this class
of models   should be controlled in turn by the string scale $M_s$. As we said, in models with a low string scale one needs dimensions
transverse to the SM branes to become very large, to understand why $M_p\gg M_s$.  This cannot be achieved in a purely toroidal model 
with intersecting D6-branes, because then some or all of the gauge couplings become negligibly small
\footnote{There are intersecting D5-brane toroidal models at singularities in which one can safely take two transverse directions
very large still maintaining gauge coupling constants of observed size \cite{IMRD5}. We have preferred not to use these models here  as
examples since their description is slightly more technical. For those the geometrical  parameters $\xi$ mentioned below are of order one.}. Still it is feasible in 
other generic CY compactifications in which the SM D6-branes wrap only a local region of the compactification in which volumes 
are not large. We can make a heuristic estimate of the relationship between  the decay constant $f$ and the string scale as follows.
The kinetic term of the axion field $\eta$ may be written as
\beq 
\frac { M_p^2}{8\pi (S+S^*)} \partial_\mu (\frac {\eta}{8\pi^2} )  \partial^\mu (\frac {\eta}{8\pi^2} )
\eeq
where $Re\,S$ is the scalar partner of $a_0$. One can estimate the value of $Re\,S$ by recalling that the gauge coupling associated to
$SU(3)$ is approximately given by
\beq
(S+S^*)\ \simeq \ \frac {g_3^2}{2\pi} \ =\ \frac {g_s V_\Pi ^{-1}}{2\pi M_s^3} \ ,
\eeq
where $V_\Pi$ is the volume of the 3-cycle wrapped by the D6's associated to the $SU(3)$ group, $g_s$ is the string coupling 
and $M_s= (\alpha')^{-1/2}$ is the string scale.   Taking into account that
\beq 
M_p^2 \ =\ \frac {8 V_6}{g_s^2(2\pi)^6\alpha«^4}
\eeq
one obtains for the axion decay constant 
\beq 
f\ \simeq \ \frac {M_s}{(2\pi)^{13/2}} \xi \ \simeq \  \xi  \times 10^{-5}M_s \ ,
\eeq
Here $\xi$ is a geometric factor, which in the toroidal case is $\xi=(V_6/V^2_\Pi)^{1/2}$, but one expects $\xi \simeq 1$ for more realistic models 
in which the SM is localised in a CY region with volumes not too different from the string scale.  So one expects the decay constant $f$ to be well below the string scale. Let us however emphasise that the precise evaluation of $f$ within a given realistic model would require details about
the geometry of the compactification and how all the moduli are fixed. The message here is that  one has $f\ll M_s$ and hence 
a value of  $f\simeq 10^2-10^3$ GeV is not in contradiction with the LHC bounds yielding  $M_s \gtrsim  7$ TeV.  Let us finally note that the fact that 
$f$ is well below $M_s$ in localised brane models can be shown in other contexts, see the Type IIB example below.

Up to now we have not discussed how the axion $a_0$ gets a large mass, possibly of order 750 GeV. As we said above, it would be very attractive if the 
axion $a_0$ here discussed had a monodromy structure so that a 4-form exists which can induce a non-trivial potential and an axion mass.
In Type IIA orientifolds of the type in our example such couplings do exist. The 10D action contains couplings of the form (see e.g.\cite{BOOK})
\beq
S_{IIA} \ \propto \  - \int _{10D} ( |F_4|^2  \ +\  F_4\wedge H_3\wedge C_3 \ +\ ...) \ .
\eeq
Here $F_4$ is the field strength of the Type IIA 3-form $C_{\mu \nu \rho }$ with indices in Minkowski space and $H_3$ is the  (quantized)  flux associated to
the Neveu-Schwarz 2-form $B_2$ with indices in compact dimensions. Expanding $C_3$ and $H_3$ in terms of harmonic 3-forms basis
$(\alpha_j,\beta^i)$ \cite{IIAflux}
\beq
H_3 \ =\  \sum_i  \  H_i \beta^i   \ ,\   C_3\ =\ \sum_j  \eta_j  \a_j  \   \ ,\    \int_{CY} \alpha_j \wedge \beta^i= \delta^i_j  \ ,
\eeq
one obtains the structure
\beq
S_{a_i} \ \propto \  - \ (|F_4|^2 \ +\ F_4 H^ia_i ) \ .
\eeq
Using the equations of motion for $F_4$ and allowing the latter to have a quantized value one has an $\eta_i=a_i/(2\pi f_i)$ axion potential of the form
\beq
V\ =\ \sigma f^4 |  n_0 - \sum_ih_i\eta_i|^2  \ ,  \  n_0,h_i \in {\bf Z} \ ,
\eeq
where on dimensional grounds we have set the overall scale of order $f^4$, with $\sigma$ a model dependent fudge factor.
Upon  discrete shifts  $a_i\rightarrow a_i+2\pi f_i$, the potential remains invariant with a shift $n_0\rightarrow n_0+h_i$. 
We see that in the case of the axion $a_0$ considered in the above example, one finds an axion mass given by
\beq
m^2_{a_0} \ =\ \sigma f^2 h_0^2 \ .
\eeq
The precise value is controlled by the model-dependent geometrical factor $\sigma$ and the quantized $NS$ flux $h_0$.
However one expects  that axion masses of order $f$ to be natural.  

Both facts, having axions coupling to QCD and hypercharge, not getting a Stuckelberg mass,  but getting  a mass  instead through fluxes is not a
particular property of Type IIA orientifolds but seems to be present more generally in string compactifications with a low string scale.
Let us briefly describe how similar ingredients  seem  to arise in a class of Type IIB orientifold models with a compact 
CY manifold with  {\it swiss cheese}  structure \cite{fernando}. These are Type IIB models with a large volume structure (see \cite{Maharana:2012tu} for an
introduction and
references). In the simplest canonical models of this class one has  two complex Kahler moduli  $T_b$ and $T_s$ with real parts
$\tau_b,\tau_s$ and $\tau_b\gg \tau_s$. The Kahler potential has a structure
\beq
\kappa^2_4 K \ =\ -2\log(\tau_b^{3/2} \ -\ \tau_s^{3/2})\ \simeq \ -3\log(\tau_b) \ +\ 2 \left( \frac {\tau_s}{\tau_b}\right)^{3/2}
\eeq
Let us assume that the SM is realised through a local set of intersecting D7-branes in which the  three branes corresponding 
to QCD are wrapping a 4-cycle with volume parametrized by the small modulus $\tau_s$. 
The rest of the SM gauge interactions will be assumed to reside in other (intersecting) 4-cycles. 
Thus we have a $U(3)$ gauge kinetic
function $f_{U(3)}=T_s/2\pi$. The hypercharge generator  here will contain the $U(1)$ inside $U(3)$, so that 
the axion $Im\,T_s$ will couple both to QCD and hypercharge, but in principle not to $SU(2)$. We will also assume that
 $Im\,T_s$ does not get a mass from a Stuckelberg coupling, something which is a model dependent issue.
  Then it is an easy exercise to 
compute what is the
size of the axion decay constant. One finds
\beq
f =\ \left(\frac {3}{32\pi}\right)^{1/2} \frac {1}{(\tau_s\tau_b^3)^{1/4}} \frac {M_p}{8\pi^2}\ =\ 
\left(\frac {9\alpha_{U(3)}}{\pi^2g_s}\right)^{1/4} \frac {M_s}{16\pi^2} \ .
\eeq
Taking $\alpha_{U(3)}\simeq g_s\simeq  0.1$, one obtains $f\simeq 5\times 10^{-3}M_s$. Thus again the 
axion decay constant is well below the string scale, as in the Type IIA case discussed above.

In this case the mass of the axion will not arise from standard closed string fluxes, which in Type IIB 
orientifolds only give masses to the complex structure and complex dilaton fields. However non-geometric fluxes  \cite{nongeometric}
may give rise to such masses in a way quite similar to the Type IIA axions discussed above. In particular, in
SUSY toroidal settings a superpotential term proportional to $W_{ng}=h_iT_i$ is created \cite{nongeometric}. This is  mirror to the Type IIA one
$W=h_iU_i$ which originates the mass term for the axions $a_i$ above.

%D-branes at singus?? 

Let us close by noting that this axion state appearing in this class of string models is expected to come
along with extra $Z$'s which could also be detected at LHC  \cite{Ghilencea:2002da,Ghilencea:2002by,Anchordoqui:2012wt}. Indeed,  there are 3 linear combinations of 
the $U(1)_{a,b,c,d}$'s which are orthogonal to hypercharge and become massive by combining with axions
(or their 2-form duals) as in eq.(\ref{BFS}). There is in fact a $4\times 4$ mass matrix for the $U(1)$'s given by \cite{Ghilencea:2002da}
\beq
(M^2)_{\alpha \beta} \ =\ \frac {M_s^2}{4\pi} g_\alpha g_\beta  \sum_i c_i^\alpha c_i^\beta \ , \ \alpha,\beta=a,b,c,d   \ ,
\eeq
where 
$g_\alpha,g_\beta$ are the corresponding $U(1)$ coupling constants, and the $c_i^\alpha$ 
are the integer coefficients appearing in eq.(\ref{BFS}). This matrix has a zero eigenvalue $M_1=0$ corresponding 
to hypercharge. There are other three massive eigenvalues $M_2,M_3,M_4$. As pointed out in refs.
 \cite{Ghilencea:2002da,Ghilencea:2002by,Anchordoqui:2012wt}
 in the toroidal
setting described in the Appendix \cite{IMR} and others based on intersecting D5-branes \cite{IMRD5}, one eigenvalue
is always above the string scale, but the other two are most often lighter, with one of them  $M_3$ in the range
$0.15 \  M_s  \leq M_3 \leq 0.32 \ M_s$ \cite{Ghilencea:2002da}.  So beyond a $750$ GeV axion one could find at LHC an extra $Z'$ 
of this class before  reaching the string threshold. Present bounds on $Z'$ from LHC stand around a region $1.5-3$ TeV
depending on the decay products \cite{ATLAS:2015nsi,Khachatryan:2015dcf,ATLAS:dilepton13,CMS:dilepton13}. However it was shown in ref.~\cite{Martin-Lozano:2015vva} that $Z'$s lighter than the maximum bound value for the mass of 3 TeV could evade those searches by a reduction of their couplings. So it could be that in the forthcoming LHC run such $Z$'s in the 1.5-5 TeV 
region could be produced.

%&&&&&&&&&&&&&&&&&&&&&&&&&&&&&&&&&
\section{Conclusions}
%&&&&&&&&&&&&&&&&&&&&&&&&&&&&&&&&&
In the present paper we have analysed whether the hints for a  750 GeV resonance recently obtained by  ATLAS and CMS experiments
 could be explained in terms of a heavy string axion in a scheme with low scale string theory.  We have shown how in
such models with a string scale $M_s\simeq 7-10^4$ TeV,  there naturally appear  massive pseudoscalar fields with
axion-like couplings both to gluons and photons, but not to $W$'s. We have exemplified this in the context of intersecting brane
models in Type IIA orientifolds, in which the SM gauge bosons reside on D6-branes and quarks and leptons live
at the intersection.

  Interestingly, in the simplest toroidal examples there is a unique axion-like scalar $a_0$ with these properties,
with all other axions in the theory becoming massive through a Green-Schwarz mechanism.  We have shown how this axion has the
correct couplings and a typical axion decay constant  $f\simeq 10^{-2}-10^{-5}\ M_s$.

Standard axions are notorious for being 
perturbatively massless,  due to their characteristic shift symmetry, and so it seems hard to understand how an axion field
could get a mass as large as 750 GeV.   We show that  the solution to this puzzle is automatic if the axion is a 
monodromy axion, of the type recently discussed in the context of string monodromy inflation \cite{mono,KS,msu} and, more recently,
relaxion models \cite{relaxion}. Monodromy axions may have a non-trivial scalar potential, and hence a mass, as long as not only the
axion transforms under the discrete  gauge shift symmetry  $a_0\rightarrow a_0+2\pi f$, but the potential parameters do.
The structure is better described in terms of quantized Minkowski 4-forms \cite{KS,BIV}.  In this paper we have shown how the 
axion $a_0$  in intersecting brane Type IIA models has the correct couplings and scalar potential of a monodromy axion 
in the presence of NS 3-form fluxes.  This behaviour is not exceptional and we have also discussed how the same type of
consisten axions and couplings arise in other string settings like large volume Type IIB orientifolds.

We have analysed the phenomenological prospects of such a heavy  axion (we call it {\it megaxion}) in describing  the 
hinted resonance at 750 GeV. Describing the observed production and decay rates set constraints on the plane of  axion couplings
$\kappa_g$ and $\kappa_\gamma$, fig.~(\ref{fig:constraints}).  If we further impose an axion width of order $45$ GeV as hinted by ATLAS, 
one is restricted to two regions, one  with  $\kappa_g \simeq 10^{-5}, \kappa_\gamma\simeq 5\times 10^{-3}$ ${\rm GeV}^{-1}$ and the
other with $\kappa_g\simeq  10^{-3}, \kappa_\gamma\simeq 10^{-5}$ ${\rm GeV}^{-1}$. However the second possibility is excluded if we
impose the limits from dijet searches in the 8 TeV run.  The allowed region implies values for the string axion decay width
$f/g_s\simeq 10^{-2}-10^{-3}$ GeV and $f/g_\gamma \simeq 1-10^2$ GeV.  If the preliminary experimental evidence is
confirmed, these values will constraint specific low scale string models.

If the hint of a 750 GeV boson at LHC is confirmed, it would probably imply,  in one way or the other, 
 a revolution in our understanding of what lies beyond the Standard Model. We have explored here the 
 possibility  that  this boson is identified with an axion-like state from a low scale string theory.  This type of
 axion with the correct couplings and a large mass appears naturally in the context of  string models with 
 the SM living at intersecting branes.  If that identification was  correct, there would be good options to
 further observe at least one extra $Z$' at LHC before reaching the string threshold. We are looking forward to the 
 analysis of the 2016 ATLAS and CMS data for a confirmation or not of this tantalising 750 GeV state.

 \vspace{1.5cm}

\bigskip

%\newpage

\section*{Acknowledgments}

We thank  F. Marchesano, A. Uranga   and I. Valenzuela  for useful discussions. 
This work is partially supported by the grants  FPA2012-32828 from the MINECO, the ERC Advanced Grant SPLE under contract ERC-2012-ADG-20120216-320421 and the grant SEV-2012-0249 of the ``Centro de Excelencia Severo Ochoa" Programme. V.M.L. would like to thank the support of the Consolider-Ingenio 2010 programme under grant MULTIDARK CSD2009-00064 and the Spanish MICINN under Grant No. FPA2012-34694.

\newpage

\appendix
\section{ The SM at Intersecting D6-branes}
\label{apendiceIMR}

The general structure of intersecting D6-brane models involve 4 sets of  D6-branes in a Type IIA orientifold (see \cite{IMR,BOOK,interev})
There is a stack  a) with 3 D6-branes carrying gauge group $U(3)_a$, including QCD and a $U(1)_a$ ; a stack b) with 2 branes and gauge
group $U(2)_b$, containing  the EW $SU(2)$ and a $U(1)_b$; a stack c) with 1 brane, yielding a $U(1)_c$ which is proportional to the 
Cartan generator of a (would be) gauge group $SU(2)_R$ of left-right symmetric models; and a stack d) with gauge group $U(1)_d$,
which is proportional to the gauged lepton number.  Being an orientifold, there is an orientifold  $Z_2$ symmetry so that one has to include 
another set of 3+2+1+1 D6-branes denoted $a^*,b^*, c^*, d^*$ , which are the orientifold mirrors of the former. The sets of D6-branes intersect at points
in the 6 compact dimensions and at the intersection localised chiral fermions appear. The number of generations is given by the times
a given pair of D6-branes intersect.  The intersection numbers are chosen so that the obtained fermion spectrum is that of the SM.
The chiral fermion spectrum and the charges of each of them under the 4 $U(1)$'s is shown in table (\ref{tabpssm})
\begin{table}[htb] \footnotesize
\renewcommand{\arraystretch}{1.25}
\begin{center}
\begin{tabular}{|c|c|c|c|c|c|c|c|}
\hline Intersection &
 Matter fields  &   &  $Q_a$  & $Q_b $ & $Q_c $ & $Q_d$  & Y \\
\hline\hline (ab) & $Q_L$ &  $(3,2)$ & 1  & -1 & 0 & 0 & 1/6 \\
\hline (ab*) & $q_L$   &  $2( 3,2)$ &  1  & 1  & 0  & 0  & 1/6 \\
\hline (ac) & $U_R$   &  $3( {\bar 3},1)$ &  -1  & 0  & 1  & 0 & -2/3 \\
\hline (ac*) & $D_R$   &  $3( {\bar 3},1)$ &  -1  & 0  & -1  & 0 & 1/3 \\
\hline (bd*) & $ L$    &  $3(1,2)$ &  0   & -1   & 0  & -1 & -1/2  \\
\hline (cd) & $E_R$   &  $3(1,1)$ &  0  & 0  & -1  & 1  & 1   \\
\hline (cd*) & $N_R$   &  $3(1,1)$ &  0  & 0  & 1  & 1  & 0 \\
\hline \end{tabular}
\end{center} \caption{ Standard model spectrum and $U(1)$ charges 
\label{tabpssm} }
\end{table}
The hypercharge is given by the linear combination of $U(1)$ charges
\beq
Q_Y \ =\ \frac {1}{6}Q_a \ -\ \frac {1}{2} Q_c \ +\ \frac {1}{2}Q_d \ .
\label{hyperapp}
\eeq
This general structure may be obtained for a variety of compact CY orientifold compactification.
The simplest example is obtained in toroidal compactifications in which the 6 extra dimensions have a
$T_1^2\times T_2^2\times T_3^2$ geometry, which is  what we describe below. However one may also obtain this structure 
in more general conformal field theory orientifolds as in ref.\cite{bert}.

Let us now briefly review the  toroidal case. Each D6-brane contains Minkowski space and a 3-cycle volume $\Pi_3$ in
compact dimensions. In these  toroidal examples the 3-cycles are obtained by  each D6 wrapping once each of the 3 $T_i^2$.
In each torus $T_i^2$  each brane wraps $n_i$ times along the $x_i$ and $m_i$ times around the $y_i$ direction. Thus
each 3-cycle is denoted by the set of 6 integers $(n_1,m_1)(n_2,m_2)(n_3,m_3)$. One can then check that the intersection number of
two stacks of branes $\alpha,\beta$ is given by
\beq
I_{\alpha \beta} \ =\ \Pi_{i=1,2,3} (n_i^\alpha m_i^\beta - n_i^\beta m_i^\alpha) \ .
\eeq
It was shown in \cite{IMR} that the most general choice of wrapping numbers  $(n_i,m_i)$ yielding just the chiral fermion 
content of the SM  with three generations is given by those in table (\ref{solution}).
\begin{table}[htb] \footnotesize
\renewcommand{\arraystretch}{2.5}
\begin{center}
\begin{tabular}{|c||c|c|c|}
\hline
 $N_i$    &  $(n_i^1,m_i^1)$  &  $(n_i^2,m_i^2)$   & $(n_i^3,m_i^3)$ \\
\hline\hline $N_a=3$ & $(1/\beta ^1,0)$  &  $(n_a^2,\epsilon \beta^2)$ &
 $(1/\rho ,  1/2)$  \\
\hline $N_b=2$ &   $(n_b^1,-\epsilon \beta^1)$    &  $ (1/\beta^2,0)$  &  
$(1,3\rho /2)$   \\
\hline $N_c=1$ & $(n_c^1,3\rho \epsilon \beta^1)$  & 
 $(1/\beta^2,0)$  & $(0,1)$  \\
\hline $N_d=1$ &   $(1/\beta^1,0)$    &  $(n_d^2,-\beta^2\epsilon/\rho )$  &  
$(1, 3\rho /2)$   \\
\hline \end{tabular}
\end{center} \caption{ D6-brane wrapping numbers giving rise to a SM spectrum.
The general solutions 
 are parametrized by a phase $\epsilon =\pm1$, the NS background
on the first two tori $\beta^i=1-b^i=1,1/2$, four integers
$n_a^2,n_b^1,n_c^1,n_d^2$ and a parameter $\rho=1,1/3$.
\label{solution} }
\end{table}
In order to obtain the correct hypercharge massless $U(1)$ 
those wrapping parameters have to verify the extra constraint
\beq
n_c^1\ =\ \frac {\beta^2}{2\beta^1}(n_a^2+3\rho n_d^2) \ .
\eeq
In addition to the above SM sector there are also closed string moduli, complex structure
and axionic fields. In particular, there are  4 axion fields $\eta_i$, $i=0,1,2,3$ from the RR sector
of the theory.  Their dual 2-forms $B_2^i$ have couplings to the $U(1)$ field strengths given by
\beqa
B_2^1 &\wedge & \ {{-2\epsilon \beta^1}\over { \beta^2 } }F^{b} \nonumber \\
B_2^2 &\wedge  & \ \frac{(\epsilon  \beta^2 )}
{\rho \beta^1}(3F^a\ -\  F^{d})
\nonumber \\
 B_2^3 &\wedge  &  \ {1\over {2 \beta^2 } }
(\frac{3\beta^2 n_a^2}{\beta^1} F^a\ +\ 6\rho n_b^1F^{b} \ +\ 
2n_c^1F^c \ +\ \frac{3\rho  \beta^2 n_d^2}{\beta^1} F^d) \
\label{bfs}
\eeqa
whereas the $B_2^0$ RR field has no couplings to the $F_j$, because
$\Pi_\alpha m_j^\alpha=0$ for all the branes. Thus the axion $\eta_0$ remains massless, as mentioned in the main text.
The dual scalars $\eta^i$ 
have couplings:
\beqa
 \eta^1\ & (\frac{\epsilon \beta^2}{2\beta^1})& (F^a\wedge F^a\ -\ 3F^d\wedge
F^d)\nonumber \\
 \eta^2\ &  (\frac{3\rho\epsilon\beta^1}{2\beta^2}) & (-F^b\wedge
F^b\ +\
2F^c\wedge
F^c)\nonumber
\\
  &  \eta^0     &  
(\frac{n_a^2}{\rho \beta^1} F^a\wedge F^a
\ +\ \frac{n_b^1}{\beta^2} F^b\wedge F^b \ +\ \frac{n_d^2}{\beta^1} 
F^d\wedge F^d) \ .
\label{cff}
\eeqa
The last equation here yields the coupling in eq.(\ref{c0coup}). In particular
there is a large class of models  with  $n_b^1=0$ in which the axion $\eta_0$ does not 
couple to the $W$ gauge bosons, as stated in the main text.

Let us finally mention that in this class of models,  the proton is stable because baryon number is a gauged (though anomalous)
gauge symmetry, which perturbatively forbids proton decay. Baryon number violation may only appear from gauge string instanton
effects, which are generically exponentially suppressed.

\end{document}